\documentclass[twocolumn]{aastex631}
\usepackage{showyourwork}

\usepackage{multirow}

\newcommand{\jax}{\texttt{JAX}\xspace}
\newcommand{\ripple}{\texttt{ripple}\xspace}
\newcommand{\lalsuite}{\texttt{lalsuite}\xspace}
\newcommand{\flowMC}{\texttt{flowMC}\xspace}

\begin{document}

\title{\ripple: Differentiable and Hardware-Accelerated Waveforms for Gravitational Wave Data Analysis}

\newcommand{\JHU}{\affiliation{William H. Miller III Department of Physics and Astronomy, Johns Hopkins University, Baltimore, Maryland 21218, USA}} 
\newcommand{\OKC}{\affiliation{The Oskar Klein Centre, Department of Physics, Stockholm University, AlbaNova, SE-106 91 Stockholm, Sweden}} 
\newcommand{\NORDITA}{\affiliation{Nordic Institute for Theoretical Physics (NORDITA), 106 91 Stockholm, Sweden}}
\newcommand{\Ciela}{\affiliation{Ciela -- Computation and Astrophysical Data Analysis Institute, Montréal, Quebec, Canada}}
\newcommand{\UdeM}{\affiliation{Département de Physique, Université de Montréal, 1375 Avenue Thérèse-Lavoie-Roux, Montréal, QC H2V 0B3, Canada}} 
\newcommand{\Mila}{\affiliation{Mila -- Quebec AI Institute, 6666 St-Urbain, \#200, Montreal, QC, H2S 3H1}} 
\newcommand{\WI}{\affiliation{Weinberg Institute, University of Texas at Austin, Austin, TX 78712, USA}}
\newcommand{\CGP}{\affiliation{Center for Gravitational Physics, University of Texas at Austin, Austin, TX 78712, USA}}
\newcommand{\CCA}{\affiliation{Center for Computational Astrophysics, Flatiron Institute, New York, NY 10010, USA}}
\newcommand{\CUHK}{\affiliation{Department of Physics, The Chinese University of Hong Kong, Shatin, N.T., Hong Kong}}

\author{Thomas D.~P.~Edwards} \JHU \OKC \NORDITA
\author{Kaze W.~K.~Wong} \CCA
\author{Kelvin K.~H.~Lam} \CCA \CUHK

\author{Adam Coogan} \Ciela \UdeM \Mila
\author{Daniel Foreman-Mackey} \CCA
\author{Maximiliano Isi} \CCA
\author{Aaron Zimmerman} \WI

\begin{abstract}
    We propose the use of automatic differentiation through the programming framework \jax for accelerating a variety of analysis tasks throughout gravitational wave (GW) science.
    Firstly, we demonstrate that complete waveforms which cover the inspiral, merger, and ringdown of binary black holes (i.e.~IMRPhenomD) can be written in \jax and demonstrate that the serial evaluation speed of the waveform (and its derivative) is similar to the \lalsuite implementation in \texttt{C}.
    Moreover, \jax allows for GPU-accelerated waveform calls which can be over an order of magnitude faster than serial evaluation on a CPU.
    We then focus on three applications where efficient and differentiable waveforms are essential.
    Firstly, we demonstrate how gradient descent can be used to optimize the $\sim 200$ coefficients that are used to calibrate the waveform model. 
    In particular, we demonstrate that the typical \textit{match} with numerical relativity waveforms can be improved by more than 50\% without any additional overhead.
    Secondly, we show that Fisher forecasting calculations can be sped up by $\sim 100\times$ (on a CPU) with no loss in accuracy.
    This increased speed makes population forecasting substantially simpler.
    Finally, we show that gradient-based samplers like Hamiltonian Monte Carlo lead to significantly reduced autocorrelation values when compared to traditional Monte Carlo methods.
    Since differentiable waveforms have substantial advantages for a variety of tasks throughout GW science, we propose that waveform developers use \jax to build new waveforms moving forward.
    Our waveform code, \ripple, can be found at \href{https://github.com/tedwards2412/ripple}{github.com/tedwards2412/ripple}, and will continue to be updated with new waveforms as they are implemented.

\end{abstract}

\section{Introduction}
\label{sec:intro}

The discovery of gravitational waves (GWs)~\citep{LIGOScientific:2016aoc} from inspiraling and merging compact objects (COs) has revolutionized our understanding of both fundamental physics and astronomy~\citep[e.g.][]{LIGOScientific:2021djp,LIGOScientific:2021sio,LIGOScientific:2021psn}.
Although the data volumes from GW detectors such as Advanced LIGO~\citep{LIGOScientific:2014pky} and Virgo~\citep{VIRGO:2014yos} are relatively small, analyzing the data is a computationally demanding task.
In addition, this computational cost will substantially increase when next generation detectors come online~\citep{Maggiore:2019uih, Reitze:2019iox, Evans:2021gyd}.

The complexity begins even before data taking, since GW searches using the matched-filtering technique~\citep{Owen:1998dk, Owen:1995tm} require the generation of large banks of template waveforms.
Once potential candidates are found, parameter estimation (PE) is performed to extract the detailed source properties of each event~\citep{Christensen:2022bxb, 2020MNRAS.493.3132S, Ashton:2018jfp, Romero-Shaw:2020owr, Veitch:2014wba, Biwer:2018osg, kombine, 10.1093/mnras/stv2422}.
For binary black holes with non-aligned spins, this requires a Markov Chain Monte Carlo (MCMC) on a 15 dimensional parameter space.
More general binary inspirals, such as those involving neutron stars, can lead to a significant increase in dimension.
Beyond these simple scenarios, more complex waveform models with additional parameters may be used to include calibration errors~\citep{Farr:2014,Vitale:2020gvb}, model the tides of neutron stars~\citep{LIGOScientific:2018hze}, and test for deviations from General Relativity~\citep{Arun:2006yw, Agathos:2013upa, Yunes:2016jcc, LIGOScientific:2016lio, LIGOScientific:2020tif, Krishnendu:2021fga}.
Finally, using the results of PE, population synthesis models constrain the progenitors systems from which the black holes we see merging today began their journey~\citep{LIGOScientific:2020kqk, LIGOScientific:2021psn, Wong:2022flg}.
Overall, GW data analysis therefore requires significant computation.
In this paper, we will argue that differentiable waveforms (and more generally differentiable pipelines) can play a significant role in alleviating this computational demand.

Derivatives are ubiquitously useful throughout data analysis tasks.
For instance, during PE, derivative information can be used to guide an optimizer towards higher-likelihood values (e.g.\,using gradient descent~\citep{2016arXiv160904747R}) or allow a sampler to rapidly explore parameter space (e.g.\, using Hamiltonian Monte Carlo (HMC)~\citep{2011hmcm.book..113N,2017arXiv170102434B}).
Gradients are particularly valuable for high dimensional spaces.
Unfortunately, in the field of GW data analysis, analytic derivatives of the necessary quantities (such as the likelihood) have historically required a significant amount of work to obtain~\citep{Keppel:2013kia}.
With waveforms only increasing in complexity, calculating analytic derivatives will become ever more tricky.
Numerical derivatives also suffer from accuracy issues stemming from rounding or truncation errors.
However, recent progress in automatic differentiation (AD) has shown promise in allowing general, fast derivative calculations for gravitational waveforms, with applications to constructing template banks~\citep{Coogan:2022qxs} and computing the Fisher information for forecasting~\citep{Iacovelli:2022bbs, Iacovelli:2022mbg}.

Automatic differentiation is a family of methods used to compute machine-precision derivatives with little computational overhead. 
AD's recent ascendance is primarily driven by its use in machine learning, particularly for derivative computations of neural networks which use gradient descent during training.
The core idea of AD is that any mathematical function can be broken down into a small set of basic operations, each with a known differentiation rule.\footnote{
    Of course, non-differentiable functions exist and care must be taken when treating these special cases.
    }
The full derivative can then be constructed using the chain rule.
There are now a variety of AD implementations, most notably in deep learning frameworks such as \texttt{pytorch}~\citep{pytorch} and \texttt{tensorflow}~\citep{tensorflow2015-whitepaper}.
More general frameworks exist in \texttt{julia}~\citep{zygote, forwarddiff}, although \texttt{julia}'s limited use in GW analysis software precludes its general use.
Here we make use of \jax~\citep{jax2018github} due to its easy integration with \texttt{python} libraries, seamless support for running code on different hardware accelerators, such as graphical processing units (GPUs), and its just-in-time (JIT) compiler, which can substantially accelerate code.

There are a variety of gravitational waveforms currently used in analysis pipelines.
They are generally structured into different families, the most common of which are: the effective-one-body (EOB)~\citep{Damour:2008yg,Buonanno:2005xu, Buonanno:2000ef, Buonanno:1998gg, Damour:2000we,Ossokine:2020kjp,Nagar:2021gss,Ramos-Buades:2021adz,Albertini:2021tbt, Cotesta:2020qhw, Bohe:2016gbl}, the phenomenological inspiral-merger-ringerdown (IMRPhenom) \citep{Husa:2015iqa,Khan:2015jqa,Hannam:2013oca,Pratten:2020ceb, Smith:2016qas, Pratten:2020fqn, Garcia-Quiros:2020qpx}, and numerical relativity surrogate (NRsurrogate)~\citep{Blackman:2017pcm,Varma:2018mmi,Varma:2019csw}.
Of these, the IMRPhenom family serves as a natural starting point for an AD implementation in \jax.
Models like the non-precessing IMRPhenomD~\citep{Khan:2015jqa} model studied here and the precessing, higher-mode model IMRPhenomXPHM~\citep{Pratten:2020fqn, Pratten:2020ceb} are written in the frequency domain using closed-form expressions.
This makes a \jax implementation that complies with the constraints of JIT compilation simple.
NRsurrogate models, which interpolate directly over waveforms produced by numerical relativity (NR) simulations, are in principle also straightfoward to implement in \jax.
EOB waveforms on the other hand are produced by evolving the dynamics of an effective one body Hamiltonian, and are therefore more difficult to implement in \jax.
For EOB waveforms, frequency-domain reduced-order models may be a convenient target~\citep[e.g.]{Cotesta:2020qhw}.

In this paper we argue that differentiable waveforms will be a vital component for the future of GW data analysis.
In addition, we present \ripple, a small GW \texttt{python} package which, at the time of writing, includes a \jax implementation of the IMRPhenomD waveform and will be continually updated with new waveforms as they are implemented.
The remainder of this paper is structured as follows. 
In Sec.~\ref{sec:waveforms} we discuss the differentiable IMRPhenomD waveform implemented in \ripple and perform some benchmarks to demonstrate its speed and accuracy. 
In Sec.~\ref{sec:applications} we discuss three distinct applications using differentiable waveforms. 
Firstly, we illustrate how the fit coefficients that form part of the IMRPhenom waveform models could be improved by high dimensional fitting enabled by a differentiable waveform. 
Secondly, we implement differentiable detector response functions and show that the speed of Fisher matrix calculations can be substantially accelerated using AD.
Finally, we run an illustrative injection example using Hamiltonian Monte Carlo to demonstrate that the autocorrelation of derivative samplers is substantially smaller than that of traditional MCMCs.
Practically, this translates into much faster PE with no sacrifice in accuracy.
The associated code can be found at \ripple~\citep{ripple}.

\section{Differentiable Waveforms}
\label{sec:waveforms}

A variety of waveform families have been developed to accurately model the GW emission from COs~\citep{Schmidtreview}. 
When the COs are relatively well separated, the dynamics of the system can be well approximated by a post-Newtonian expansion.
However, close to merger, NR simulations are required to accurately model the binary.
Unfortunately, these numerical simulations are computationally expensive and cannot be run in conjunction with data analysis.
Approximate, phenomenological waveforms have therefore been constructed to enable relatively fast waveform generation at sufficient accuracy.

As mentioned in the previous section, the three major waveform families that have been developed to date are:  EOB,  NRsurrogate, and IMRPhenom.
EOB waveforms require one to model the binary system using a Hamiltonian and are typically slow to evaluate, whereas IMRPhenom waveforms are constructed with simple closed-form expressions.
IMRPhenom waveforms are therefore ideally-suited for AD, especially using \jax. 
In this paper we focus on the aligned-spin, circular-orbit model IMRPhenomD~\citep{Husa:2015iqa, Khan:2015jqa}.

The implementation of IMRPhenomD in \lalsuite is in C, and therefore needs to be rewritten natively into \texttt{python} to be compatible with \jax.
We have re-written IMRPhenomD from scratch using a combination of pure \texttt{python} and \jax derivatives.
In addition, we have restructured the code for readability and evaluation speed as well as exposing the internal fitting coefficients to the user (which we will use later in Sec.~\ref{sec:applications}).

To demonstrate our implementation of IMRPhenomD is faithful to the \lalsuite implementation, we start by defining the noise weighted inner product:
\begin{equation}
    \label{eq:inner_prod}
    \left(h_1|h_2\right) \equiv 4 \, \mathrm{Re} \int^{\infty}_{0} \mathrm{d} f \, \frac{ h^*_1(f) h_2(f)}{S_n(f)}\, ,
\end{equation}
where $S_n$ is the (one-sided) noise power spectral density (PSD) and $h_1$ and $h_2$ are the frequency domain waveforms which are to be compared.
We can then normalize the inner product through
\begin{equation}
    \left[h_1|h_2\right] = \frac{\left(h_1|h_2\right)}{\sqrt{\left(h_1|h_1\right)\left(h_2|h_2\right)}}\, .
\end{equation}
Now we are ready to define the \textit{match} which is given by 
\begin{equation}
    \mathrm{m}(h_1|h_2) \equiv \max_{\Delta t_c,\, \Delta \phi_c} \left[h_1 \, \middle| \, h_2 \right]\,
\end{equation}
where $\Delta t_c$ and $\Delta \phi_c$ are, respectively, the differences in time and phase of coalescence between the two waveforms.
Finally, we can define the mismatch as 
\begin{equation}
    \label{eq:mismatch_1}
	\mathcal{M}(h_1|h_2) \equiv 1- \mathrm{m}(h_1|h_2)\,.
\end{equation}

\begin{figure}[t]
    \script{random_matches.py}
    \begin{centering}
        \includegraphics[width=\linewidth]{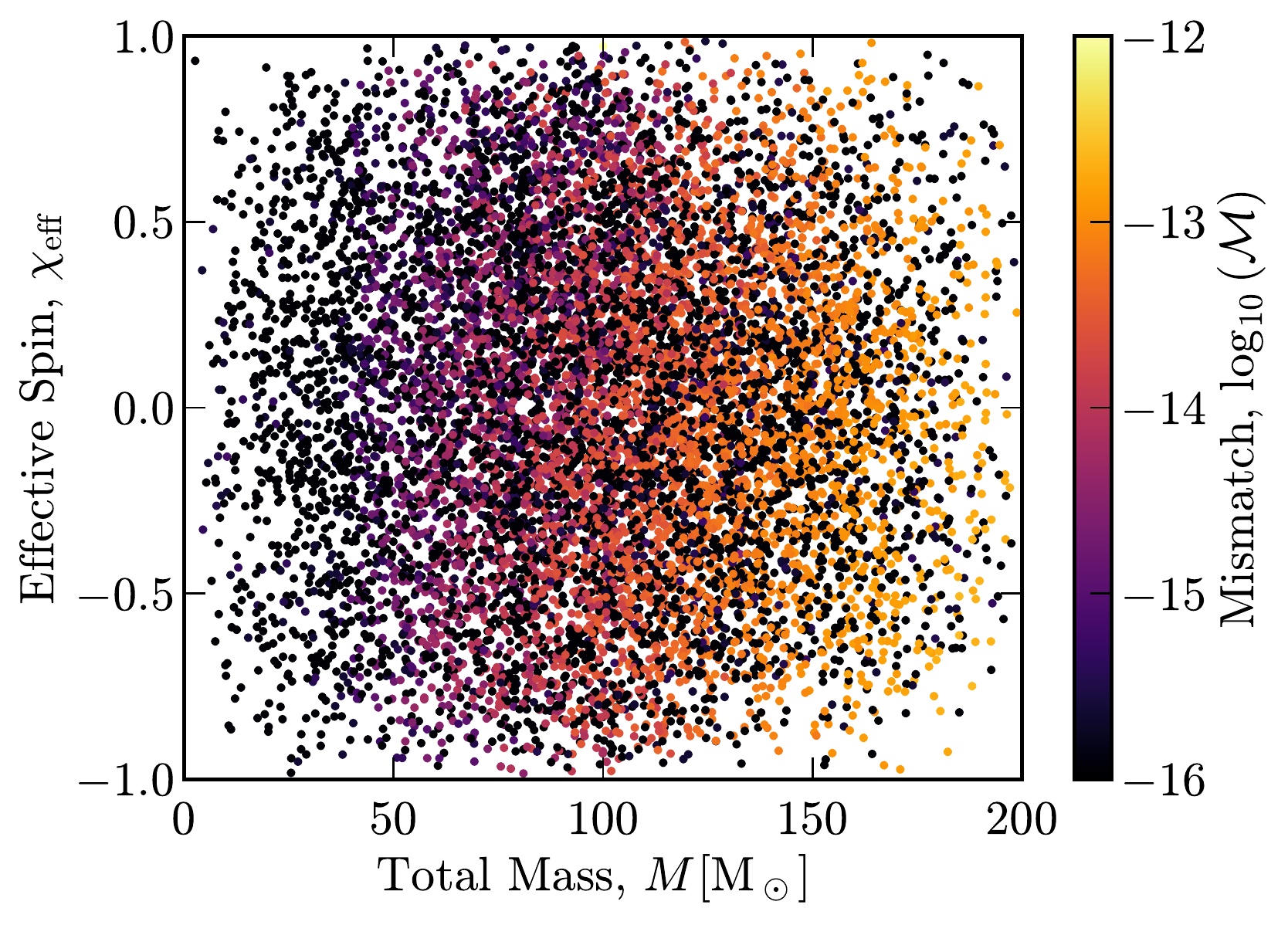}
        \caption{
            Match between the \ripple and \lalsuite implementations of the IMRPhenomD waveform as a function of total mass and effective spin. 
            Points which give $\mathcal{M}=0$ (up to numerical precision) are plotted as the lowest value in the colorbar.
            It is clear that the \ripple waveform matches the \lalsuite implementation many of orders of magnitude more than necessary for all data analysis tasks across the entire parameter space.
        }
        \label{fig:match}
    \end{centering}
\end{figure}

Since the match is a measure of the difference between two waveforms, we can use it to demonstrate that the implementation of IMRPhenomD in \ripple accurately matches the \lalsuite implementation. For this comparison, we use the $S_n$  presented in GWTC-2~\citep{LIGOScientific:2020ibl} from the Livingston detector (the most sensitive of the detectors).\footnote{
    \url{https://dcc.ligo.org/LIGO-P2000251/public}.
    }
This is shown in Fig.~\ref{fig:match}, where we have calculated the $\mathrm{m}(h_1|h_2)$ across the entire parameter space.\footnote{
    Specifically, we use $10^4$ points varying component masses $m_1$, $m_2$ and spin parameters $\chi_1$, $\chi_2$ in the ranges $m_{1,2} = (1,100)\,M_{\odot}$ and $\chi_{1,2} = (-1,1)$. In addition, we evaulated the waveforms on a frequency grid from $32\,$Hz to $1024\,$Hz with frequency spacing $\Delta f = 0.0125 \,$Hz.
}
Here, $h_1$ corresponds to the \ripple waveform implementation and $h_2$ is the \lalsuite implementation, evaluated at the same point in parameter space.
From Fig.~\ref{fig:match}, it is clear that the \ripple waveform matches with the \lalsuite waveform close to numerical precision across the entire parameter space.
In fact, the scale in Fig.~\ref{fig:match} is clipped such that points which give $\mathcal{M}=0$ (up to numerical precision) are plotted as the lowest value in the colorbar.

Note that in General Relativity (GR) the total mass $M = m_1 + m_2$ of a binary black hole simply serves as an overall scale for the system, so that the frequency evolution can be trivially rescaled.
The total mass only impacts the match because the chosen PSD and frequency limits fix a reference scale.
If we instead use a flat PSD in Eq.~\eqref{eq:inner_prod} and rescale the frequency grid to units of $Mf$, all dependence with $M$ seen in Fig.~\ref{fig:match} vanishes. 
This figure instead illustrates a more realistic PSD where at low masses the waveform signal-to-noise is dominated by the inspiral but at high masses it is dominated by the merger.

There remains some slight deviation between the two waveform implementations at high total mass. 
This is partly due to the fact that cubic interpolators, which are used within IMRPhenomD to calculate the ringdown and damping frequencies, are not currently supported in \jax. 
Instead, we initially use \texttt{scipy}'s cubic interpolator to create a fine grid of $5\times10^5$ values, which we then linearly interpolate during waveform generation.
Unfortunately, we cannot make the initial cubic interpolation arbitrarily fine as this would add additional computational overhead when loading the data during waveform evaluation.
Additionally, we use a Fourier transform to efficiently maximize over $\Delta t_c$ which will inherently come with some inaccuracy.
This inaccuracy will primarily affect the mismatch at higher total masses where the waveform is dominated by the merger.
Note however, that the differences are well below the accuracy requirements of the waveforms and will have no noticeable effect for realistic data analysis tasks.

For maximum utility, a waveform needs to be fast to evaluate.
Fortunately, the IMRPhenom waveforms are constructed from simple closed-form expressions which are computationally efficient.
Even though the \lalsuite implementation of IMRPhenomD is written in \texttt{C}, the serial evaluation of the waveform in \ripple is comparably fast. 
Benchmarking on a MacBook Pro with an M1 Max Apple Silicon processor, we find that a single waveform evaluation takes $\sim 0.5\,\mathrm{ms}$ for \lalsuite (interfaced with python) and $\sim 0.4\,\mathrm{ms}$ for \ripple.\footnote{
    For this benchmark we used $24\,\mathrm{Hz}$ and $512\,\mathrm{Hz}$ for the lowest and highest frequencies respectively. 
    In addition, we used a frequency spacing of $0.2\,\mathrm{Hz}$. 
    We performed this benchmark by evaluating the waveform $10^{4}$ times and taking the average evaluation time.
}
Although similar on a CPU benchmark, \jax has a few key advantages. 
First, its ability to JIT compile allows for significant performace gains (the above benchmark is already JIT compiled).
Second, automatic vectorization can be achieved using the \texttt{vmap} function.
Using \texttt{vmap} and performing the same benchmark as described above reduces the average evaluation speed to $\sim 0.14\,\mathrm{ms}$.
Finally, \jax can natively on a GPU which allows for highly parallellized waveform evaluations.

Again, performing the same benchmark as above on a NVIDIA Quadro 6000, we find that on average waveform evaluations take $\sim 0.02\,\mathrm{ms}$, over an of magnitude faster than serial CPU evaluation.
Generalizing \lalsuite waveforms to run on a GPU would be a significant undertaking.

One of the primary aims of this paper is that waveform derivatives will also be highly valuable to data analysis tasks. 
AD provides two big advantages when it comes to evaluating derivatives compared to numerical differentiation.
First, the accuracy of derivatives from AD are significantly more stable than finite-difference methods.
In particular, finite differences suffer from both rounding and truncation errors, meaning that the user is required to \textit{tune} the width over which the difference is taken.
On the other hand, AD produces machine-precision derivatives with no tuning.
Second, AD scales favorably with the dimensionality of the function.
In particular, for every input dimension added, $D$, one would need to evaluate the function at least $2D$ times to calculate finite difference derivatives for all input parameters.
For reverse-mode AD, one only needs two function calls to evaluate the derivative of all input parameters, regardless of dimension.\footnote{
    Note that although the number of function calls is small, reverse-mode AD does add memory overhead.
    We've not found this to be limiting in any of the situations tested so far.
}
For a review of AD methods see~\cite{2018arXiv181105031M}.
Since the parameter space of GWs in GR has $\mathcal{O}(10)$ dimensions, the speed of derivative evaluation is less crucial than the stability.
However, this might change for waveforms in beyond-GR models, waveforms involving equation of state parameters for neutron stars, and models which account for calibration and waveform uncertainties.
In these cases many more parameters can be added.

Overall, we have demonstrated that the IMRPhenom waveform family is ideally suited for AD.
Moreover, we have shown that our implementation of IMRPhenomD in \ripple is accurate and quick to evaluate, especially when hardware acceleration is available.
In the next section, we will discuss a variety of potential use cases of differentiable waveforms.

\section{Applications}
\label{sec:applications}

Here, we illustrate how three core tasks in GW science can be substantially improved through the use of differentiable waveforms.
In this paper we primarily look at toy examples, leaving more careful analyses to future work. 
The three tasks discussed here cover a wide range of GW science, starting with waveform development all the way to Fisher forecasting and PE.

\subsection{Fine-Tuning Waveform Coefficients}
\label{subsec:coeffs}

Having an accurate waveform model is essential for many data analysis tasks throughout
GW science.
While waveforms generated using NR simulations are in principle the highest fidelity signal model, they are too computationally expensive to be used in any practical data analysis tasks.
The community therefore utilizes waveform ``approximants'' (such as those discussed in the introduction) which can be evaluated much faster and in regions not covered by numerical simulations, such as extremes of parameter space or earlier phases of the binary inspiral.

Waveform approximants generally have free coefficients which are calibrated
to NR waveforms to achieve high accuracy.
In the case of IMRPhenomD, there are 209 fitting coefficients used to capture the separate behavior of the amplitude and phase as a function of the mass ratio and spins.

Any inaccuracy in obtaining the fitting coefficients
leads to a misrepresentation of the NR waveform, which can translate to systematic
error in downstream data analysis tasks. For example, sufficiently large
systematic errors in the waveform would cause the recovered source parameters to be biased
in the case of PE.

Previously in the construction of IMRPhenomD \citep{Khan:2015jqa}, waveform
coefficients for the amplitude and phase were fitted independently.
Furthermore, IMRPhenomD is divided into three fitting segments: inspiral; merger; and ringdown.
Each of these segments has their own set of fitting coefficients.
After obtaining the fitting coefficient for individual segments, they are then ``stitched'' together such that both the phase and amplitude are continuous in the first derivative.
The process of stitching introduces some additional inaccuracy in the waveform model, as the connections affect the originally fitted segments.

The coefficients of the original IMRPhenomD implementation are tuned in subsets of parameters instead of all together. 
This means the tuning process ignores the correlation between different subsets of parameters, so the provided best-fit solution may not be the global optimum. 
We therefore aim to improve of the accuracy of the waveform by jointly fitting all coefficients at once.

In general, optimization problems in a high dimensional space benefit from having access to the gradient of the objective function. 
Since we can differentiate through the entire waveform model against the fitting parameters, one can use gradient descent to more efficiently find the local best fitting parameters.

The first step is to define a loss function that measures the goodness-of-fit of the current waveform coefficients.
Here, we choose it to be the mismatch between the NR waveform and the approximant waveform:
\begin{equation}
    \label{eq:mismatch}
	\mathcal{M}(\lambda)=1-\mathrm{m}(h_{\theta}^{\mathrm{IMR}}(\lambda)|h_{\theta}^{\mathrm{NR}}(\lambda))\,,
\end{equation}
where $\lambda$ is a vector of the fitting coefficients, $h_{\theta}^{\mathrm{IMR}}$ is
the waveform generated by IMRPhenomD, and $h_{\theta}^{\mathrm{NR}}$ is the waveform
generated by the NR simulation. 
Given the loss function, we use gradient descent to update the fitting
coefficients:
\begin{equation}
	\lambda\leftarrow\lambda-\alpha\nabla\mathcal{M}\,, 
\end{equation}
where $\alpha$ is the learning rate. We set $\alpha$ to be $10^{-6}$. 

To generalize the loss function to a collection of waveforms we use the average of the mismatch of individual waveforms, given by
\begin{equation}
	\mathcal{L}=\frac{1}{N}\sum_{i=1}^{N}\mathcal{M}_i,
    \label{eq:multi_loss}
\end{equation}
where $\mathcal{M}_i$ is the mismatch of an individual training waveform and $N$ is the total number of training waveforms used in the optimization.
This optimization is more difficult since we are now applying the same set of coefficients to waveforms with different intrinsic parameters, such as mass ratio and spins. 
From Eq.~\eqref{eq:multi_loss}, we can see the averaging between waveforms with different intrinsic parameters implies there are trade-offs in performance for different regions of the intrinsic parameter space. 
Additionally, this means our best-fit points will generally depend on the distribution of training waveforms across the parameter space.

To evaluate Eq.~\eqref{eq:multi_loss} we use a flat PSD and a frequency array which is scaled by the total mass.\footnote{
    In particular, we use $M f_l = 2.5 \times 10^{-3}$, $M f_u = 0.2$, and $M\Delta f = 2.5 \times 10^{-6}$ for the dimensionless lower, upper, and frequency spacing respectively. We fix $M$ to be $50\,\mathrm{M}_\odot$ throughout this section.
}
Since the NR waveforms do not have a total mass associated with them, we assign a fixed total mass to compare with the \ripple waveform.

For training, we use the publically available subset (11 waveforms) of the 19 waveforms used in the original IMRPhenomD paper~\citep{Khan:2015jqa}. 
These 11 waveforms are taken from the SXS catalog~\citep{Boyle:2019kee}.
We then run gradient descent, as described above, for $3\times 10^4$ steps. 
Figure~\ref{fig:loss_compare} shows the relative error (against a test NR waveform; see below) of the original and optimized waveform as a function of dimensionless frequency.
The vertical dashed lines indicate the stitching points for the phase i.e. when the inspiral is joined onto the merger.
We can see the error of optimized waveform is lower than that of the original waveform for most of the domain.
In particular, the error in the merger region is decreased by half while other regions also show good improvement in accuracy. 

\begin{figure}[t]
	\script{loss_compare.py}
	\begin{centering}
		\includegraphics[width=\linewidth]{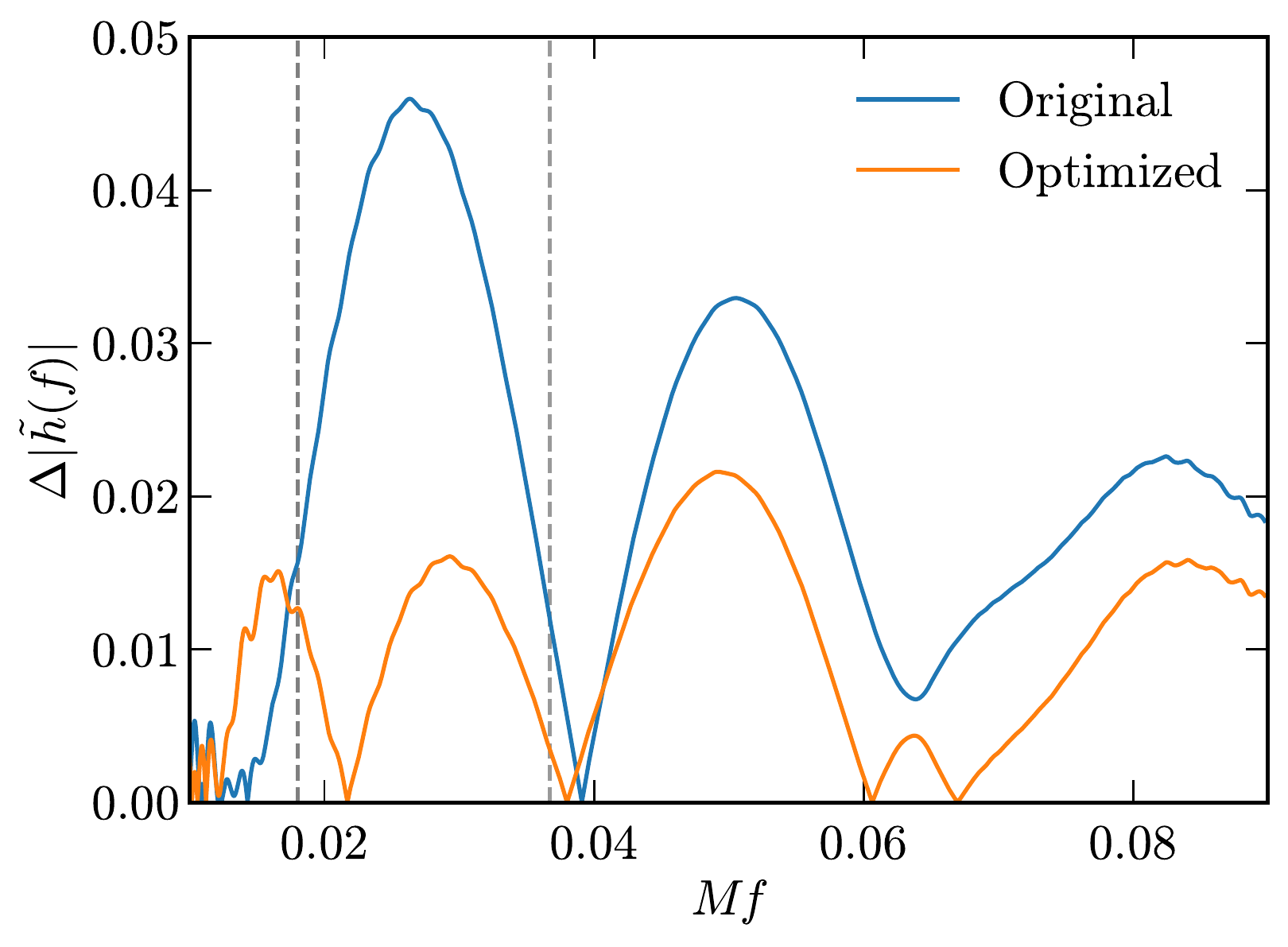}
		\caption{
			Relative error between the numerical relativity and IMRPhenomD waveform amplitudes. 
            The blue line shows the relative error of the IMRPhenomD model with the original model coefficients where as the orange line uses the IMRPhenomD model after optimization using gradient descent.
            Interestingly, the error is reduced across the frequency range, demonstrating that high dimensional optimization is useful for all parts of the waveform.
        }
		\label{fig:loss_compare}
	\end{centering}
\end{figure}

\begin{figure}[t]
	\script{mismatch_hist.py}
	\begin{centering}
		\includegraphics[width=\linewidth]{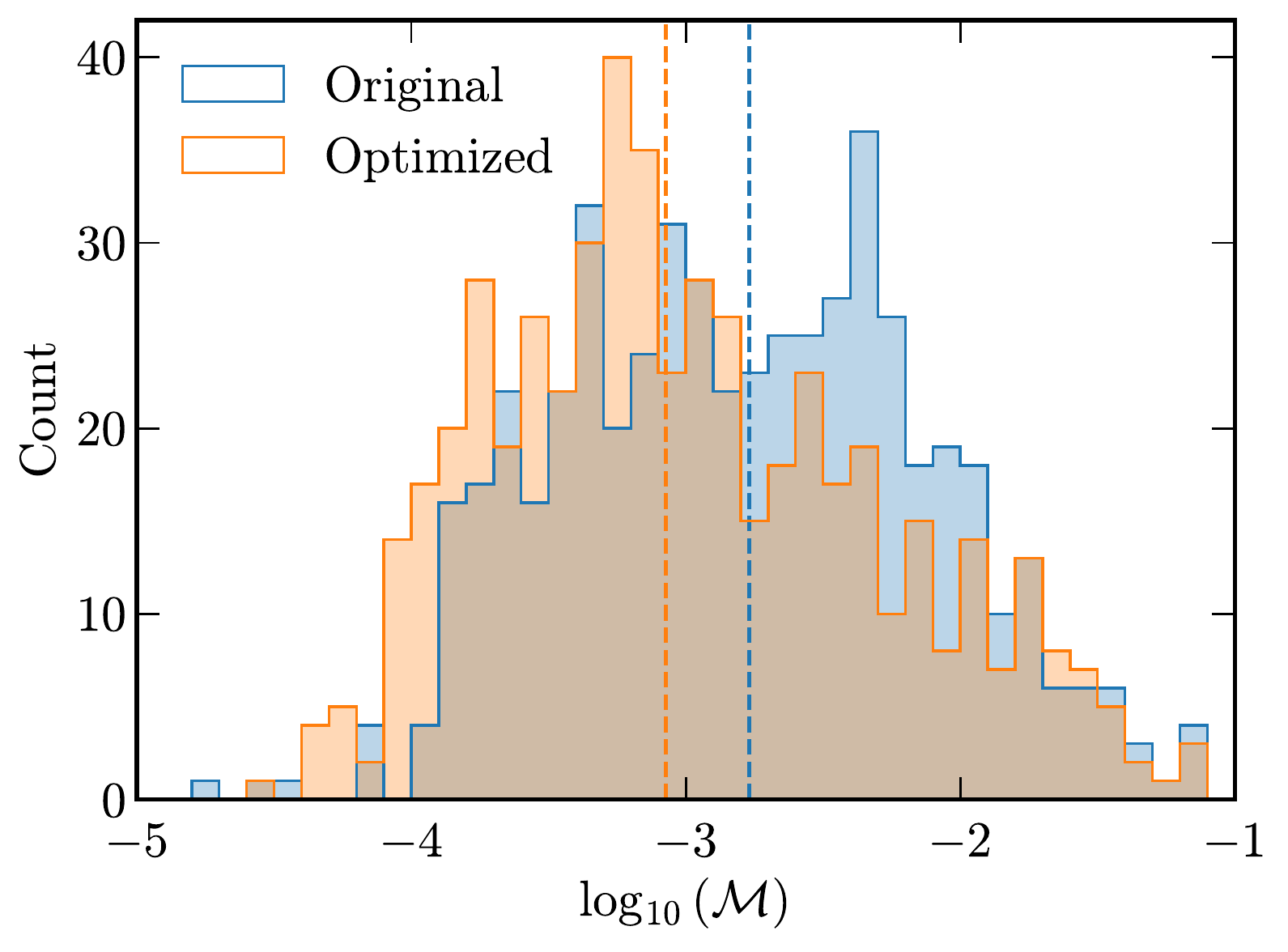}
		\caption{
			Distribution of 536 log mismatches (see Eq.~\eqref{eq:mismatch}) for the original (blue) and optimized (orange) IMRPhenomD. 
            Overall, the optimized distribution shifts to lower mismatches and therefore a better waveform model. 
            More quantitatively, the median mismatch (shown as vertical dashed lines) is reduced by $\sim50$\%. 
		}
		\label{fig:mismatch_hist}
	\end{centering}
\end{figure}
 
In Fig.~\ref{fig:mismatch_hist}, we show the distribution of log mismatches for a set of test waveforms.
In particular, we use 536 waveforms from the SXS catalog (we simply choose all waveforms with aligned spins, $\chi_{x,y}<5\times 10^{-3}$, and eccentricity $<2\times 10^{-3}$)~\citep{Boyle:2019kee}.
One can see that the distribution of mismatches after optimization is generally shifted to lower mismatch compared to the original waveform.
In particular, the peak of the original waveform distribution has moved by nearly an order of magnitude, indicating that our AD-assisted optimization procedure provides an improved implementation of the model.

While we focused on \texttt{IMRPhenomD} here, the ability to apply AD to the calibration parameters may assist in other approaches to calibration, such as that used for the aligned-spin EOB model in~\citet{Bohe:2016gbl}.
There, MCMC methods were used in a two-step procedure to optimize the calibration parameters. 
Derivative information, if implemented for EOB waveforms, could allow for the application of other sampling methods such as HMC, or possibly optimization over the entire set of NR waveforms at once.

\subsection{Fisher Forecasting}
\label{subsec:fisher}

\begin{figure}[t!]
    \script{sky_localization.py}
    \centering
    \includegraphics[width=\linewidth]{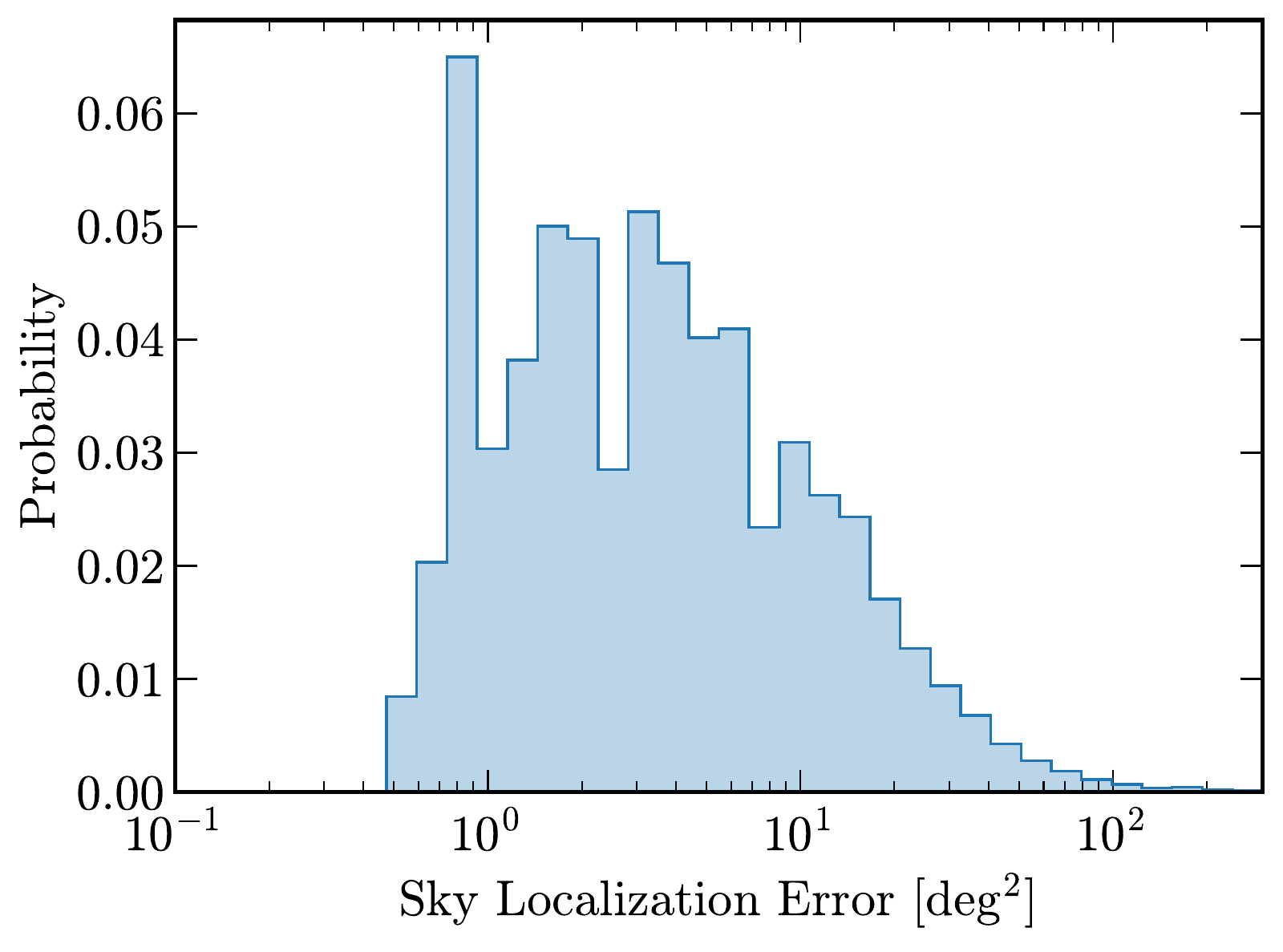}
    \caption{
        Distribution of Fisher information sky localization error for a population of nearby binaries.
        Automatic differentiation (AD) was used to compute the derivatives in Eq.~\eqref{eq:fisher}.
        As emphasized in the text, after JIT compilation, each error calculation was around 100 times faster than a similar forecasting code \texttt{GWbench}~\citep{Borhanian:2020ypi}.
    }
    \label{fig:sky_localization}
\end{figure}

Forecasting the sensitivity of future experiments is a routine task in GW science. 
Due to its theoretical simplicity and evaluation speed, the Fisher matrix formalism~\citep{Cornish:2010kf} is commonly deployed to estimate how well a binary system's parameters could be measured.
The Fisher matrix approach is built around the assumption of a Gaussian likelihood~\citep{Vallisneri:2007ev}.
Although in practice this assumption is often violated for realistic detector noise, the results obtained using a Fisher analysis can provide quick and useful diagnostics in evaluating sensitivities for a variety of models and detector configurations.

Computing the Fisher matrix requires one to evaluate derivatives of the likelihood, which in turn involves derivatives of the waveform model and detector projection functions.
AD is therefore perfectly suited for computing Fisher matrices accurately and efficiently. 
Forecasting with Fisher matrices for third generation detectors has already been extensively explored in~\citep{Iacovelli:2022bbs, Iacovelli:2022mbg}.
Here we purely want to illustrate the simplicity and speed of AD for forecasting rather than providing new physics insights.
We therefore consider a simple, three-detector setup corresponding to the two LIGO detectors in addition to Virgo.

The Fisher information matrix for a single detector is typically given by 
\begin{equation}
    \label{eq:fisher}
    \mathcal{I}^{k}_{ij} = (\partial_i h^k | \partial_j h^k) \, ,
\end{equation}
where $k$ indicates the detector, $\partial_i = \partial/\partial \theta_i$, and $h^k$ is the strain measured by the detector which is given by,
\begin{equation}
    h^k(\theta) = F^k_+(\phi) h_{+}(\Xi) + F^k_\times(\phi) h_{\times}(\Xi) \, .
\end{equation}
Note that here we have separated out the extrinsic ($\phi$) and intrinsic ($\Xi$) variables as well as introducing the detector projection functions for the plus and cross polarizations as $F^k_+$ and $F^k_\times$ respectively.
Since we are considering a three detector setup we simply add the Fisher matrices from the individual detectors to get the combined Fisher matrix:
\begin{equation}
    \mathcal{I}_{ij} =  \mathcal{I}^{\mathrm{Hanford}}_{ij} + \mathcal{I}^{\mathrm{Livingston}}_{ij} + \mathcal{I}^{\mathrm{Virgo}}_{ij}   \, .
\end{equation}
Finally, we invert the Fisher matrix to calculate the covariance matrix,
which provides forecasted measurement errors and parameter covariances for a signal with parameters $\theta$ observed by the given detector network in the high signal-to-noise limit.

\begin{table}[t]
    \centering
    \begin{tabular}{c|c}
    \hline\hline

    $m_1, m_2$ & $\boldsymbol{U}[20, 60]\, \mathrm{M}_\odot$ \\ \hline
    $\chi_1, \chi_2$ & $\boldsymbol{U}[-0.8, 0.8]$ \\ \hline
    $D$ (Uniform in Volume) & $[600, 900]$\,Mpc \\ \hline
    $t_c$ & 0.0 \\ \hline
    $\phi_c$ & 0.0 \\ \hline
    Inclincation Angle, $\cos(\iota)$ & $\boldsymbol{U}[-1, 1]$ \\ \hline
    Polarization Angle, $\psi$ & $\boldsymbol{U}[0, 2\pi]$ \\ \hline
    Right Ascension, $\alpha$ & $\boldsymbol{U}[0, 2\pi]$ \\ \hline
    Declination, $\sin(\delta)$ & $\boldsymbol{U}[-1, 1]$ \\ 
    \hline\hline
    \end{tabular}
    \caption{Priors for the 11 dimensional parameter space used for the Fisher forecasting population analysis in Sec.~\ref{subsec:fisher}. 
    $\boldsymbol{U}$ indicates a uniform distribution between the two variables in the brackets. 
    The distance prior ensures that the binaries are uniformly distirbuted in volume.
    }
    \label{tab:priors}
\end{table}

To illustrate the computational speed of computing Fisher matrices with AD, we consider a population of binaries and compute the sky localization error following Eq.~(28) in \cite{Iacovelli:2022bbs, Iacovelli:2022mbg}.
Since the Fisher matrix approach is known to have both theoretical issues as well as numerical instabilities for low signal-to-noise events, we restrict our population to only nearby systems.
A full list of the distributions used to generate the various parameters in our population are given in Tab.~\ref{tab:priors}. 
Additionally, we use $20\,$Hz, $1024\,$Hz, and $16\,$s for the minimum frequency, sampling frequency, and sample length.
Our noise curves correspond to the design PSDs for LIGO Hanford, LIGO Livingston (\textsc{SimNoisePSDaLIGOZeroDetHighPower}) and Virgo (\textsc{SimNoisePSDAdvVirgo}).\footnote{\url{https://lscsoft.docs.ligo.org/lalsuite/lalinspiral/psds_8py_source.html}.}
The resulting population produces binaries with signal-to-noise ratios ranging from $\mathcal{O}(10-10^2)$.

The distribution of sky localization errors from a population of $10^3$ binaries can be seen in Fig.~\ref{fig:sky_localization}.
We have verified that our errors agree with a separate dedicated Fisher forecasting code~\citep{Borhanian:2020ypi} to within ~30\%.
This demonstrates that AD can be used to accurately produce population-level forecasts.

Moreover, each error calculation (including computing the Fisher matrices for each detector and the inversion process) is substantially faster.
In particular, we find that after compilation, each error calculation takes approximately half a second on a single computing core.
\texttt{GWbench}~\citep{Borhanian:2020ypi}, on the other hand, takes $\mathcal{O}$(minutes) for each Fisher calculation using the same detector setup and frequency grid.
This factor of over 100 speed up is substantial considering the fact that a single core evaluation of the \ripple waveform is similar to \lalsuite which is used by \texttt{GWbench}.
On a MacBook Pro with an M1 Max Apple Silicon processor, JIT compilation takes $\sim 270\,s$ and the full population analysis takes less than $\sim 9$ minutes.
As discussed above, performance can be further improved by utilizing hardware acceleration such as parallel GPU processing.
AD therefore represents a fast and accurate way of performing population level analyses, and should be utilized for testing the capabilities of next generation detectors.

\subsection{Derivative Based Samplers - Hamiltonian Monte Carlo}
\label{subsec:hmc}

\begin{figure*}[t]
    \script{corner_plot.py}
    \centering
    \includegraphics[width=\linewidth]{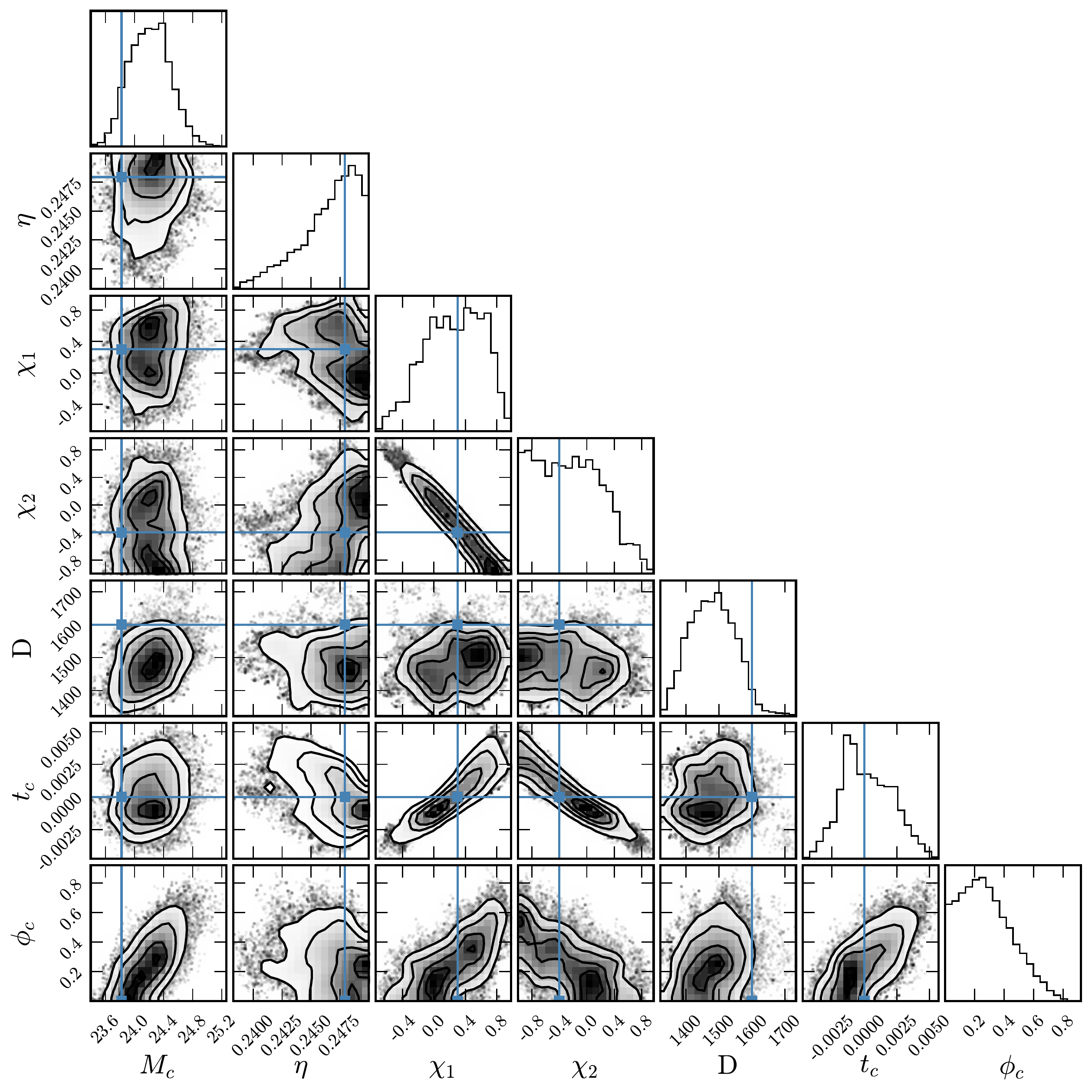}
    \caption{
        Corner plot for the posteriors from an HMC (see text for details) on simulated noise with injected signal. 
        Orange lines indicate the true values of the injection. 
        Although not fully converged, it is clear that we find posteriors consistent with the injected parameters. 
    }
    \label{fig:corner}
\end{figure*}

After the search algorithms have constructed a list of confidently detected binaries, the next step is to sample from the posterior of each sources parameter's - so called PE.
To do this, one typically uses an MCMC or nested sampler~\citep{Skilling2004, multinest, dynesty}.
Although robust, both MCMC and nested sampling are slow to converge and are known to perform poorly in high dimensional parameter spaces.
For example, sampling the 15 dimensional parameter space for a BBH system can take $\mathcal{O}(10)$ hours, while BNS systems can take up to weeks.
Dedicated fast samplers have been designed to get approximate posteriors on the sky localization to facilitate follow-up electromagnetic observations \citep[e.g.~\texttt{BAYESTAR}][]{Singer:2015ema}. 
Moreover, a number of methods have been developed to speed up PE well below the numbers quoted above~\citep{Dax:2021tsq, Islam:2022afg,Roulet:2022kot, Zackay:2018qdy, Cornish:2021lje, Canizares:2013ywa, Leslie:2021ssu}.
Nevertheless, these do not present the whole picture; fast, general PE therefore remains a key aim of GW data analysis.

A primary issue with both MCMC and nested sampling is that neither utilizes information about the likelihood's derivative and must therefore randomly walk towards areas of highest likelihood.
Derivative based samplers, on the other hand, have been shown to extrapolate well to higher dimensions although they sometimes come with their own drawbacks.
Here we simply aim to demonstrate the utility of a derivative based sampler and its efficiency on a small test problem.
In particular, we will show that the autocorrelation of a Hamiltonian Monte Carlo (HMC) sampler is significantly lower than a traditional MCMC algorithm~\citep{Bouffanais:2018hoz, 2014CQGra..31n5004P}.

For our basic example we perform an injection recovery test on a seven dimensional parameter space with the two LIGO detectors in our network.
Our noise curves correspond to the design PSDs for LIGO Hanford and Livingston (\textsc{SimNoisePSDaLIGOZeroDetHighPower}).
We use $20\,$Hz, $1024\,$Hz, and $16\,$s for the minimum frequency, sampling frequency, and sample length.
More specifically, we generate Gaussian noise consistent with the measured PSDs for each detector and then inject a BBH signal with parameters: chirp mass $M_c = 23.82\,\mathrm{M_\odot}$,  symmetric mass ratio $\eta = 0.248$, primary spin parameter $\chi_1=0.3$, secondary spin parameter $\chi_2=-0.4$, luminosity distance $D = 1.6\,\mathrm{Gpc}$, coalescence time $t_c = 0.0$, and coalescence phase $\phi_c = 0.0$.\footnote{
    The remaining parameters (inclination angle $\iota$, polarization angle $\psi$, right ascension $\alpha$, declination $\delta$) are set to $\pi/3$.
}
Using a standard Gaussian likelihood, we then run the HMC sampler implemented in \flowMC~\citep{2022arXiv221106397W} for $2\times10^5$ steps and the random walk Metropolis Hastings (RWMH) sampler~\citep{1953JChPh..21.1087M} for $1.5\times10^6$ steps (each with four randomly initialized independent chains). 
The number of steps and and mass matrix used for each example was hand tuned to give good performance for the specific sampler.
The additional steps for the GRW sampler were required to achieve a similarly converged posterior.

We note at this point that neither pure HMC nor RWMH are the most modern versions of gradient and non-gradient based samplers.
For example, for gradient based samplers one could use a No-U-Turn sampler~\citep{2011arXiv1111.4246H} or the Metropolis-adjusted Langevin algorithm~\citep{2013arXiv1309.2983X}.
Traditional MCMC methods such as nested sampling~\citep{Skilling2004, multinest, dynesty} or the Affine Invariant MCMC Ensemble sampler implemented in \texttt{emcee}~\citep{2013PASP..125..306F} will also lead to more efficient sampling of the posterior than basic RWMH.
Here we instead seek to demonstrate the simplicity with which HMC can be implemented within a differentiable pipeline.
In addition, as we discuss below, we find that a basic HMC algorithm will produce significantly more efficient sampling that RWMH, motivating further exploration of gradient-based samplers for GW PE~\citep{PEpaper}.

In Fig.~\ref{fig:corner}, the grey contours show the posterior recovered using the best chain (i.e.~one that reached the highest log-likelihood values).
The orange shows the true parameters of the injected signal.
From the one dimensional histograms along the diagonal, it is clear that we consistently recover all seven parameters apart from $\phi_c$.
This is expected since the injected binary is relatively nearby with an SNR of $\sim13$.

\begin{figure}[t]
	\script{autocorrelation.py}
	\begin{centering}
		\includegraphics[width=\linewidth]{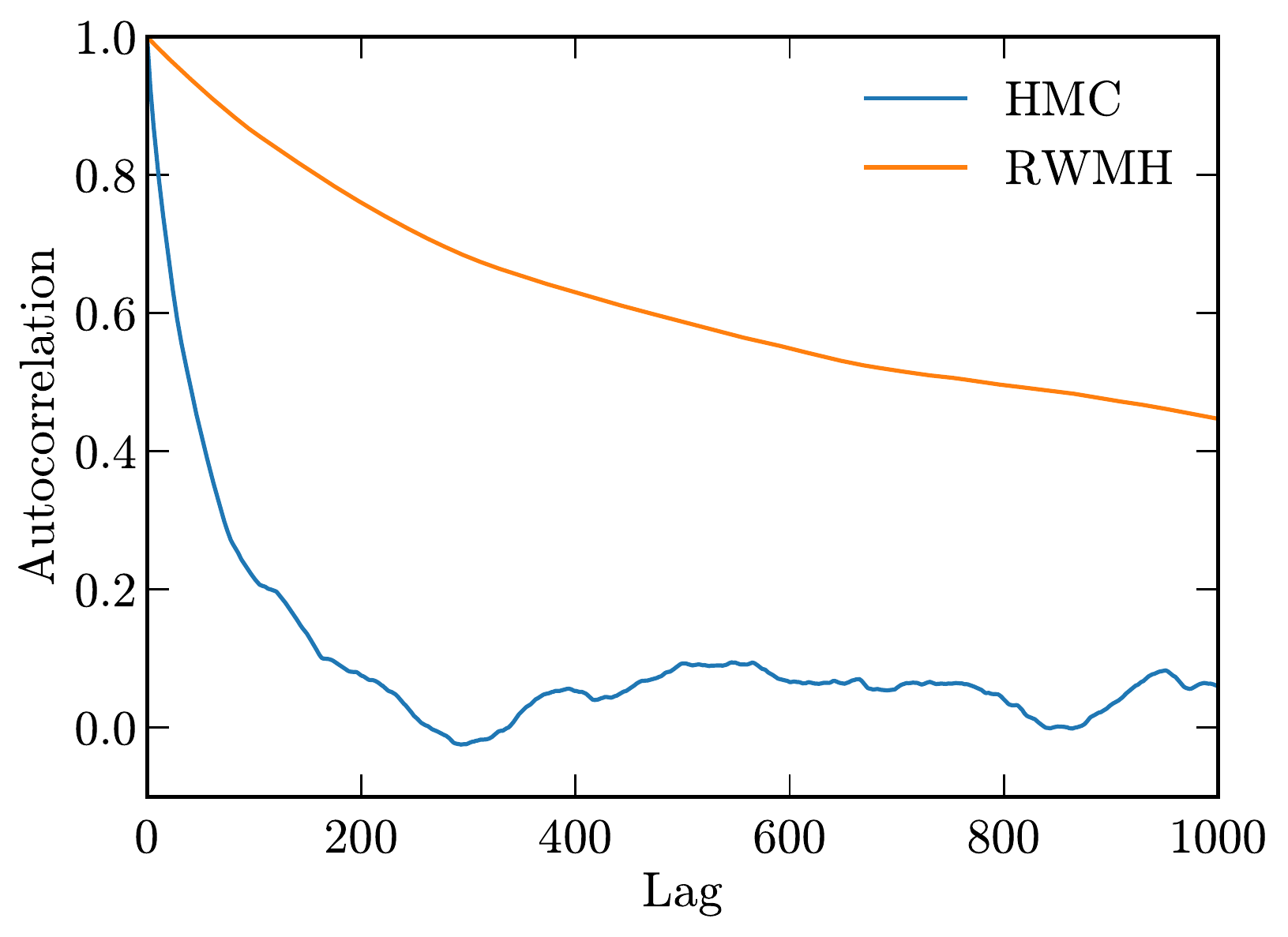}
		\caption{
			Autocorrelation as a function of lag for HMC and GRW on the same simulated data as discussed in Fig.~\ref{fig:corner}.
            The smaller autocorrelation of the HMC leads to a larger number of independent samples and therefore a faster converging Monte Carlo.
		}
		\label{fig:autocorrelation}
	\end{centering}
\end{figure}

Although further steps would be required to achieve a fully-converged posterior, these chains are sufficient to show the increased efficiency associated with HMC. 
To further illustrate this, in Fig.~\ref{fig:autocorrelation} we plot the autocorrelation as a function of lag for both the HMC and GRW samplers.
The HMC autocorrelation is substantially lower than that of the GRW.
We therefore expect gradient based samplers to converge significantly faster than typical samplers, especially in higher dimensions.
In addition, we found that the effictive number of samples~\citep{arviz_2019},\footnote{
    Computed using \textsc{arviz}~\url{https://python.arviz.org/en/stable/api/generated/arviz.ess.html}
    }
(a measure of the number of indepedent samples) is between 2 and 7 times larger for HMC across the different dimensions of the parameter space.

In a follow up paper we will demonstrate that minute scale PE can be achieved by combining normalizing flows~\citep{2022arXiv221106397W, Gabrie:2021tlu}, GPU acceleration, and a derivative based sampler~\citep{PEpaper}. 
We therefore expect \jax waveforms to be highly beneficial to future PE efforts in GW astronomy, particular for low-latency pipelines and higher dimensional analyses.

\section{Discussion and Conclusion}
\label{subsec:discussion}

In this paper we introduced and discussed the various benefits of differentiable waveforms in \jax for GW data analysis.
First, we demonstrated the speed and accuracy of our implementation of the aligned spin IMRPhenomD waveform.
In particular, we showed that it matches the \lalsuite implementation to near machine precision and can be easily parallelized on a GPU.
Parallelization on a GPU provides substantial speed increases; on a NVIDIA Quadro 6000 GPU we found that waveform evaluations are over an order of magnitude faster than serial CPU evaluations.
Second, we discussed three data analysis tasks which can all be substantially improved by utilizing derivative information of the waveform.
Although we primarily discuss toy examples in this paper, each can be extended to the full data analysis task, some of which will be shown in upcoming papers~\citep{PEpaper}.
Differentiable waveforms therefore represent a crucial advancement towards efficient GW science.

In this paper, we have primarily focussed on the IMRPhenom family of waveforms as their closed form expression is perfectly suited for a \jax implementation.
Two other waveform families are commonly used in GW data analysis: the effective-one-body (EOB) and numerical relativity surrogate (NRSurrogate). 
A differentiable NRsurrogate implementation is under development~\citep{NRSurAD}, but it currently seems difficult to implement EOB waveforms in \jax.
In particular, the evolution of the Hamiltonian required to evaluate an EOB waveform is both inherently slow to differentiate and difficult to implement in \jax. 
Since EOB methods are used to produce state-of-the-art waveforms for many applications, more work is required to see if a fast, differentiable implementation is possible.

Currently the biggest constraint to adopting differentiable waveforms is the need to rewrite the most commonly used waveforms into \jax (or pure python).
In order to showcase the benefits of differentiable waveforms as quickly as possible, at the time of writing, we have only implemented an aligned spin GW model (IMRPhenomD).
We plan on adding a variety of different waveforms to \ripple in the near future with the primary goal of reaching a \jax version of a fully precessing, higher order mode waveform such as IMRPhenomXPHM~\citep{Pratten:2020ceb}.
Ideally, future waveforms should be implemented under an AD framework such as \jax.
This would ensure that the community can easily utilize differentiability and hardware acceleration in the future.

\section{Acknowledgments}
This material is based upon work supported by NSF's LIGO Laboratory which is a major facility fully funded by the National Science Foundation.
This work was supported by collaborative visits funded by the Cosmology and Astroparticle Student and Postdoc Exchange Network (CASPEN). 
T.E.\ is supported by the Horizon Postdoctoral Fellowship.
A.Z.~is supported by NSF Grant PHY-2207594.
A.~C.\ acknowledges funding from the Schmidt Futures Foundation.
This document has been assigned preprint numbers UTWI-4-2023 and LIGO-P2300025.

\bibliography{bib}

\end{document}